\documentclass[showpacs,preprintnumbers,amsmath,amssymb,twocolumn]{revtex4-1}
\usepackage{graphicx}
\usepackage{dcolumn}
\usepackage{bm}
\usepackage{subfigure}
\usepackage{float}
\usepackage{multirow}
\usepackage{booktabs}
\usepackage{amsmath}
\usepackage{latexsym,amssymb}
\usepackage[mathscr]{eucal}
\usepackage[switch*]{lineno}
\usepackage[bookmarks=true,
   colorlinks=true,
   linkcolor=blue,
   urlcolor=blue,
   citecolor=blue,
   bookmarks=true,
   hyperindex=true
]{hyperref}

\begin{document}

\title{Quantum corrections to the thermodynamics and phase transition of a black hole surrounded by a cavity in the extended phase space}

\author{Zhong-Wen Feng\textsuperscript{1}}
\altaffiliation{Email: zwfengphy@163.com}
\author{Xia Zhou\textsuperscript{1}}
\author{Shi-Qi Zhou\textsuperscript{2}}
\author{Shu-Zheng Yang\textsuperscript{1}}
\vskip 0.5cm
\affiliation{1 School of Physics and Astronomy, China West Normal University, Nanchong, 637009, China\\
2. School of Physics and Astronomy, Sun Yat-sen University, Zhuhai 519082, China}


\begin{abstract}
In the extended phase space, we investigate the rainbow gravity-corrected thermodynamic phenomena and phase structure of the Schwarzschild black hole surrounded by a spherical cavity. The results show that rainbow gravity has a very significant effect on the thermodynamic phenomena and phase structure of the black hole. It prevents the black hole from total evaporation and leads to a remnant with a limited temperature but no mass. Additionally, we restore the  $P-V$ criticality and obtaine the critical quantities of the canonical ensemble. When the temperature or pressure is smaller than the critical quantities, the system undergoes two Hawking-Page-like phase transitions and one first-order phase transition, which never occurs in the original case. Remarkably, our findings demonstrate that the thermodynamic behavior and phase transition of the rainbow SC black hole surrounded by a cavity in the extended phase space are analogous to those of the Reissner-Nordstr\"{o}m anti-de Sitter black hole. Therefore, rainbow gravity activates the effect of electric charge and cutoff factor in the evolution of the black hole.
\end{abstract}
\keywords{Rainbow gravity; Phase transitions;  Extended phase space; Schwarzschild black hole}
\maketitle
\section{Introduction}
\label{Int}
In the past fifty years, the Hawking radiation of black holes has been among the most hottes topics in theoretical physics, which has helped researchers research the relationship between gravity, quantum mechanics, and thermodynamics \cite{cha1,cha2}. In addition, a variety of research works show that black holes not only have temperature and entropy, but also have rich phase structures and critical phenomena. In 1983, Hawking and Page \cite{cha4} demonstrated that there is a phase transition between pure thermal radiation and a stable anti-de Sitter (AdS) space. This seminal work is so important because it can be used to explain the deeper-seated relation between confinement and deconfinement phase transition of the gauge field in the AdS/CFT correspondence \cite{cha5,cha5+}. Since then, studying the phase transitions of AdS black holes has become a fascinating topic \cite{cha6,cha6+,cha7,cha8,cha9}. More interestingly, by assuming the cosmological constant $\Lambda$ of AdS spacetime as a thermodynamic pressure $P$ \cite{cha10++,cha10+,cha3}, Kubiz\v{n}\'{n}ak and Mann analyzed the critical behavior of the Reissner-Nordstr\"{o}m (R-N) AdS black hole in an extended phase space. Their results showed that the phase behavior and $P-V$ criticality of the R-N AdS black hole coincide with those of Van der Waals gas \cite{cha10}. Subsequently, the thermodynamic properties and phase transition of various AdS black holes in the extended phase space were discussed Refs.~\cite{cha11,cha12,cha13,cha14,cha14+,cha15,cha16,cha17,cha18,cha18+,cha18++,cha18+++}.

Generally, the essential reason the AdS spacetimes have such intriguing phase structures is based upon the fact that the AdS boundary condition plays the role of a reflecting surface, which allows the AdS black holes in a quasilocal thermally stable structure and makes the study of their phase behavior possible. In addition, York proposed another popular method to discuss the phase structures and critical phenomena of non-AdS spacetimes, namely, enclosing the classical black hole inside a cavity. In this way, the cavity acts as the spacetime in the same way as the AdS space. According to this approach, York found that the Schwarzschild (SC) black hole in a cavity exists the Hawking-Page-like transition, which is similar to that of the SC AdS black hole \cite{cha19}. In Refs.~\cite{cha20,cha21,cha22,cha23}, the authors confirmed that in the canonical ensemble and grand canonical ensemble, the R-N black hole in a cavity has the same phase transition behavior as the AdS case. Along the line, similar investigations have been extended to a variety of complicated spacetimes \cite{cha24,cha26,cha26+,cha27,cha27+,cha28}.

Although the aforementioned works have led to many insights into the properties of black holes, there remain some defects. On the one hand, the former scenario is not suitable for asymptotically flat black holes without the AdS term since the lack of the reflecting wall would cause the thermodynamic instability of systems. On the other hand, the latter scenario is always used to investigate the black holes in the normal phase space, in which the spacetimes background is fixed (i.e., the radius of cavity is regarded as a constant), which leads to the $P-V$ criticality and the corresponding thermodynamic phenomena of the black hole systems in the extended space are absent. To address these issues, many new research schemes have been proposed in recent years. In Ref.~\cite{cha25},  Simovic and Mann investigated the thermodynamic behavior of SC de Sitter and R-N de Sitter black holes enclosed in an isothermal cavity by treating the cosmological constant as thermodynamic pressure and found a novel pressure-dependent phase transition that never occurs in the asymptotically AdS spacetime. Promsiri \emph{et al}. \cite{cha28+} proposed the R\'{e}nyi extended phase space approach in which the nonextended parameter $\lambda$ is regarded as the thermodynamic pressure. In this way, they discussed the solid/liquid phase transition and heat engine of SC black holes \cite{cha28++}. Almost at the same time, Wang \emph{et al}. proposed a third scheme, which can be realized by redefining the effective volume as  $V = {{4\pi r_c^3} \mathord{\left/ {\vphantom {{4\pi r_c^3} 3}} \right. \kern-\nulldelimiterspace} 3}$ and its conjugate pressure as  $P =  - \left( {{{\partial {\cal E}} \mathord{\left/ {\vphantom {{\partial {\cal E}} {\partial V}}} \right. \kern-\nulldelimiterspace} {\partial V}}} \right)$, where ${r_c}$ and ${\cal E}$ are the radius and the thermal energy of a black hole system in a cavity, respectively. In Refs.~\cite{cha29,cha30,cha30+,cha30++}, they extended the phase space of SC, R-N and Gauss-Bonnet black holes in the cavity, and then investigated the thermodynamic properties and $P-V$ criticality of those black holes. The results showed that the black hole systems have the Hawking-Page-like phase transition. More importantly, in the extended phase spaces, it is found that the thermodynamic behavior of black holes in a cavity is similar to those of the AdS counterparts.

Nevertheless,  one also notes that previous works cannot reveal the evolution and phase structure of real black holes since the everyday thermodynamics of black holes obtained from the original methods will lead to the loss of information and naked singularity problems. An effective way to prevent these problems is to use quantum gravity (QG) \cite{cha31,cha32,cha32+}.  In particular,  by considering rainbow gravity (RG), which is an important model of QG, the catastrophic behavior of Hawking radiation and the information paradox of black holes can be effectively avoided \cite{cha33,cha33+,cha34,cha35}. Furthermore, in~\cite{cha37,cha37+,cha37++,cha38,cha39,cha40,chb40,chb41,cha41,cha42,cha43}, the authors found that the RG changes the picture of phase structures and critical phenomena. Consequently, due to the effect of RG, we showed that the thermodynamics and the phase transition of the SC black hole in a cavity are similar to those of the R-N AdS black hole, which is quite different from the conclusion in Refs.~\cite{cha44,cha44+,cha44++}. According to the discussions above, a question arises: what are the thermodynamics and the phase structures of the SC black hole (i.e., the SC black hole in the framework of rainbow gravity) surrounded by a cavity in the extended phase space? However, this question has never been examined. To this end, in this present paper, we extend the phase space of the rainbow SC black hole enclosed by a cavity and investigate its Hawking temperature, entropy, heat capacity, thermodynamic stability, critical behavior and phase transition. In particular, this allows us to determine the AdS counterpart of the rainbow black hole in a cavity in the extended phase space.

This paper is arranged as follows. In section 2, we review the rainbow SC black hole and its basic thermodynamic quantities. In section 3, we investigate the thermodynamic properties and stability of the rainbow SC black hole enclosed by a cavity in extended phase space. In section 4, the critical behavior and phase transition of the rainbow SC black hole in a cavity  is discussed in extended phase space. We summarize our results in section 5.

\section{Thermodynamics of the rainbow SC black hole}
\label{sec2}
In this section, we briefly introduce the rainbow functions and the thermodynamics of the rainbow SC black hole. According to the viewpoints of QG, the standard dispersion relation ${m^2} = {E^2} - {p^2}$ is no longer held near the Planck scale \cite{cha31,cha32}. It should be modified to the so-called modified dispersion relation (MDR).  In the ultraviolet limit, a general form of MDR is given by
\begin{equation}
\label{eq1}
{m^2} = {E^2}{f^2}\left( {{E \mathord{\left/ {\vphantom {E {{E_p}}}} \right. \kern-\nulldelimiterspace} {{E_p}}}} \right) - {p^2}{g^2}\left( {{E \mathord{\left/
 {\vphantom {E {{E_p}}}} \right. \kern-\nulldelimiterspace} {{E_p}}}} \right),
\end{equation}
where  $m$,  $E$ and $p$ represent the mass, energy and momentum of the test particle, respectively. The correction terms $f\left( {{E \mathord{\left/
 {\vphantom {E {{E_p}}}} \right. \kern-\nulldelimiterspace} {{E_p}}}} \right)$ and $g\left( {{E \mathord{\left/ {\vphantom {E {{E_p}}}} \right. \kern-\nulldelimiterspace} {{E_p}}}} \right)$ are called rainbow functions, which are phenomenological functions that depend on the Planck energy  ${E_p}$ \cite{cha45}. According to MDR~(\ref{eq1}),  Magueijo and Smolin proposed doubly general relativity, or rainbow gravity \cite{cha46}.  In this scenario, the Einstein's equations become ${G_{\mu \nu }}\left( {{E \mathord{\left/ {\vphantom {E {{E_p}}}} \right. \kern-\nulldelimiterspace} {{E_p}}}} \right) = 8\pi G{T_{\mu \nu }}\left( {{E \mathord{\left/ {\vphantom {E {{E_p}}}} \right. \kern-\nulldelimiterspace} {{E_p}}}} \right)$. This implies that the metric of spacetime felt by a test particle would depend on its energy. In particular, the expressions of rainbow functions are not unique, they are determined depending on different phenomenological motivations. Therefore, the background of spacetime is no longer a single metric, but a family of metrics parameterized by the ratio ${{E \mathord{\left/ {\vphantom {E {{E_p}}}} \right. \kern-\nulldelimiterspace} {{E_p}}}}$. However, in the low-energy limit $\mathop {\lim }\limits_{{E \mathord{\left/ {\vphantom {E {{E_p}}}} \right. \kern-\nulldelimiterspace} {{E_p}}} \to 0} f\left( {{E \mathord{\left/ {\vphantom {E {{E_p}}}} \right.  \kern-\nulldelimiterspace} {{E_p}}}} \right) = 1$ and  $\mathop {\lim }\limits_{{E \mathord{\left/ {\vphantom {E {{E_p}}}} \right. \kern-\nulldelimiterspace} {{E_p}}} \to 0} g\left( {{E \mathord{\left/ {\vphantom {E {{E_p}}}} \right. \kern-\nulldelimiterspace} {{E_p}}}} \right) = 1$, Eq.~(\ref{eq1}) reproduce the standard dispersion relation.

 In the high-energy regime, Amelino-Camelia \emph{et al}. \cite{cha31,cha32} constructed one of the most studied MDRs as follows:
 \begin{equation}
\label{eq2}
{m^2} = {E^2} - {p^2}\left[ {1 - \eta {{\left( {{E \mathord{\left/ {\vphantom {E {{E_p}}}} \right. \kern-\nulldelimiterspace} {{E_p}}}} \right)}^2}} \right].
\end{equation}
Comparing the Eq.~(\ref{eq1}) with Eq.~(\ref{eq2}), the specific rainbow functions reads
\begin{equation}
\label{eq3}
f\left( {{E \mathord{\left/ {\vphantom {E {{E_p}}}} \right. \kern-\nulldelimiterspace} {{E_p}}}} \right) = 1,\quad g\left( {{E \mathord{\left/
 {\vphantom {E {{E_p}}}} \right. \kern-\nulldelimiterspace} {{E_p}}}} \right) = \sqrt {1 - \eta {{\left( {{E \mathord{\left/ {\vphantom {E {{E_p}}}} \right.
 \kern-\nulldelimiterspace} {{E_p}}}} \right)}^2}},
\end{equation}
where $\eta$  is the rainbow parameter. Notably, the abovementioned functions are compatible with results from loop quantum gravity and  $\kappa$-Minkowski noncommutative spacetime. Now, according to the viewpoint of~\cite{cha46}, the time coordinates  ${\rm{d}}t$ and all spatial coordinates ${\rm{d}}{x^i}$  should be replaced by ${{{\rm{d}}t} \mathord{\left/ {\vphantom {{{\rm{d}}t} {f\left( {{E \mathord{\left/ {\vphantom {E {{E_p}}}} \right. \kern-\nulldelimiterspace} {{E_p}}}} \right)}}} \right. \kern-\nulldelimiterspace} {f\left( {{E \mathord{\left/  {\vphantom {E {{E_p}}}} \right. \kern-\nulldelimiterspace} {{E_p}}}} \right)}}$  and  ${{{\rm{d}}{x^i}} \mathord{\left/ {\vphantom {{{\rm{d}}{x^i}} {g\left( {{E \mathord{\left/ {\vphantom {E {{E_p}}}} \right. \kern-\nulldelimiterspace} {{E_p}}}} \right)}}} \right. \kern-\nulldelimiterspace} {g\left( {{E \mathord{\left/  {\vphantom {E {{E_p}}}} \right. \kern-\nulldelimiterspace} {{E_p}}}} \right)}}$, respectively. In this sense, the line element of the rainbow SC black hole can be described as follows:
\begin{align}
\label{eq4}
{\rm{d}}{s^2}{\rm{ }} =&  - \frac{{{\cal F}\left( r \right)}}{{{f^2}\left( {{E \mathord{\left/ {\vphantom {E {{E_p}}}} \right. \kern-\nulldelimiterspace} {{E_p}}}} \right)}}{\rm{d}}{t^2} + \frac{{{\rm{d}}{r^2}}}{{{\cal F}\left( r \right){g^2}\left( {{E \mathord{\left/ {\vphantom {E {{E_p}}}} \right. \kern-\nulldelimiterspace} {{E_p}}}} \right)}}
 \nonumber \\
&+\frac{{{r^2}}}{{{g^2}\left( {{E \mathord{\left/ {\vphantom {E {{E_p}}}} \right. \kern-\nulldelimiterspace} {{E_p}}}} \right)}}\left( {{\rm{d}}{\theta ^2} + {{\sin }^2}\theta {\rm{d}}{\phi ^2}} \right),
\end{align}
where$\mathcal{F}\left( r \right) = 1 - {{{r_H}} \mathord{\left/ {\vphantom {{{r_H}} r}} \right. \kern-\nulldelimiterspace} r}$ is the metric function with the event horizon ${r_H}$.

By utilizing the rainbow functions~(\ref{eq3}), the RG corrected Hawking temperature of the SC black hole gives
\begin{equation}
\label{eq5}
T_H^{{\rm{RG}}} = {T_H}\frac{{g\left( {{E \mathord{\left/ {\vphantom {E {{E_p}}}} \right. \kern-\nulldelimiterspace} {{E_p}}}} \right)}}{{f\left( {{E \mathord{\left/
 {\vphantom {E {{E_p}}}} \right. \kern-\nulldelimiterspace} {{E_p}}}} \right)}} = \frac{1}{{4\pi {r_H}}}\sqrt {1 - \eta {{\left( {\frac{E}{{{E_p}}}} \right)}^2}},
\end{equation}
where ${T_H} = {1 \mathord{\left/ {\vphantom {1 {4\pi {r_H}}}} \right. \kern-\nulldelimiterspace} {4\pi {r_H}}}$ is the Hawking temperature of the original SC black hole. Obviously, the modified Hawking temperature is energy-dependent. To eliminate the dependence of the particle energy in Eq.~(\ref{eq5}), one can follow the findings in Refs.~\cite{cha47,cha48}, the Heisenberg uncertainty principle $\Delta x\Delta p \ge 1$  holds in the framework of RG. Therefore, the momentum between the energy and the uncertainty position can be expressed as  $p = \Delta p = {1 \mathord{\left/ {\vphantom {1 {\Delta x}}} \right. \kern-\nulldelimiterspace} {\Delta x}} \sim {1 \mathord{\left/ {\vphantom {1 {{r_H}}}} \right. \kern-\nulldelimiterspace} {{r_H}}}$. By substituting the momentum uncertainty into MDR~(\ref{eq2}), the energy for the massless particle is \cite{cha48+}
\begin{equation}
\label{eq6}
E = \frac{1}{{\sqrt {\eta  + r_H^2} }},
\end{equation}
where we used the natural units  $G = c = \hbar  = {k_B} = 1$, which lead to  ${E_p} = 1$. Plugging Eq.~(\ref{eq6}) into Eq.~(\ref{eq5}), the rainbow Hawking temperature can be rewritten as
\begin{equation}
\label{eq7}
T_H^{{\rm{RG}}} = \frac{1}{{4\pi {r_H}}}\sqrt {1 - \frac{\eta }{{r_H^2 + \eta }}}.
\end{equation}
It is clear that the modified Hawking temperature returns to the original case ${T_{\rm{H}}}$  when  $\eta  = 0$. In addition, in the limit ${r_H} \to 0$, the rainbow Hawking temperature becomes finite as  ${{\sqrt {{1 \mathord{\left/ {\vphantom {1 \eta }} \right. \kern-\nulldelimiterspace} \eta }} } \mathord{\left/  {\vphantom {{\sqrt {{1 \mathord{\left/ {\vphantom {1 \eta }} \right. \kern-\nulldelimiterspace} \eta }} } {4\pi }}} \right. \kern-\nulldelimiterspace} {4\pi }}$, indicating that the effect of RG plays the role of cutoff and regularizes the standard divergent Hawking temperature. Then, according to the first law of black hole thermodynamics, the modified entropy reads
 \begin{equation}
\label{eq8}
{S^{{\rm{RG}}}} = \pi \left[ {{r_H}\sqrt {r_H^2 + \eta }  + \eta \ln \left( {{r_H} + \sqrt {r_H^2 + \eta } } \right)} \right].
\end{equation}
It is worth noting that the effect of RG gives a logarithmic correction to the modified entropy, which satisfies the requirements of quantum gravity theory \cite{cha49,cha50,cha51}. However, Eq.~(\ref{eq8}) reduces the entropy of the original SC black hole $S = \pi r_H^2$  in the limit of  $\eta  = 0$. This result implies that the effect of RG dose contributes to the quantum corrected metric.

\section{Thermodynamics of the rainbow SC black hole surrounded by a cavity in the extended phase space}
\label{sec3}
In this section, we study the thermodynamics of the rainbow SC black hole surrounded by a cavity in the extended phase space. According to the method of York, one can enclose the rainbow SC black hole in a spherical cavity to maintain thermal stability \cite{cha19}. Therefore, the black hole system can be considered as a canonical ensemble in the following discussion. Now, suppose the radius of cavity as $r_c$,  the temperature of the rainbow SC black hole in a cavity can be expressed as follows:
\begin{equation}
\label{eq9}
T{\rm{ = }}\frac{{T_H^{{\rm{RG}}}}}{{\sqrt {{\cal F}\left( {{r_c}} \right)} }}{\rm{ = }}\frac{1}{{4\pi {r_H}\sqrt {1 - \frac{{{r_H}}}{{{r_c}}}} }}\sqrt {1 - \frac{\eta }{{r_H^2 + \eta }}},
\end{equation}
in which the above equation is implemented by the blueshifted factor of the metric of the SC black hole. Next, on the basis of the Euclidean action method \cite{cha22}, the thermal energy of the rainbow SC black hole in a cavity is given by \cite{cha53}
\begin{equation}
\label{eq10}
\mathcal{E} = {r_c}\left( {1 - \sqrt {1 - \frac{{{r_H}}}{{{r_c}}}} } \right).
\end{equation}
Notably, in order to ensure that the wall of the cavity can effectively reflect the radiation from the rainbow SC black hole, the radius of the cavity is required to be larger than the radius of the black hole, that is ${r_c} \geqslant {r_H}$  (or ${{0 \leqslant {r_H}} \mathord{\left/ {\vphantom {{0 \leqslant {r_H}} {{r_c}}}} \right.  \kern-\nulldelimiterspace} {{r_c}}} \leqslant 1$). Therefore, for ${r_c} \to \infty$, the cavity is no longer able to reflect Hawking radiation, leading to the thermal instability of the rainbow SC black hole, as the temperature in the cavity decreases the original rainbow SC black hole case and the thermal energy becomes zero, which makes it impossible to study the phase behavior of the rainbow SC black hole in the cavity. In previous works \cite{cha20,cha21,cha22,cha23,cha24,cha25,cha26,cha26+,cha27,cha27+,cha28}, the radius of the cavity $r_c$ has been considered as an invariable quantity, which leads to the absence of thermodynamic volume  $V$ and pressure $P$.  To overcome this situation, Wang \emph{et al}. \cite{cha29,cha30,cha30+} regarded $r_c$ as a thermodynamic variable and defines a new thermodynamic volume for the black hole in a cavity as follows
\begin{equation}
\label{eq11}
V = \frac{4}{3}\pi r_c^3,
\end{equation}
which gives the corresponding conjugate thermodynamic pressure
\begin{equation}
\label{eq12}
P =  - \frac{{\partial {\cal E}}}{{\partial V}} = \frac{{2{r_c}\left( {1 - \sqrt {1 - \frac{{{r_H}}}{{{r_c}}}} } \right) - {r_H}}}{{8\pi r_c^3\sqrt {1 - \frac{{{r_H}}}{{{r_c}}}} }}.
\end{equation}
From Eq.~(\ref{eq11}) and Eq.~(\ref{eq12}), we restore the ``$P-V$ conjugate pair" in a new extended phase space, which is similar to the properties of AdS spacetime. Consequently, these thermodynamic quantities obtained above satisfy the first law of black hole thermodynamics ${\rm{d}}{\cal E} = T{\rm{d}}S - P{\rm{d}}V$. In the extended phase space, it is interesting to investigate the equation of state of the canonical ensemble. To this aim, one should solve $r_c$  in terms of  $T$ according to Eq.~(\ref{eq9}), then yields
\begin{equation}
\label{eq13}
{r_H} = \frac{{{r_c}}}{3} + \frac{{2{{\left( {2\pi } \right)}^{2/3}}{T^2}\left( {r_c^2 - 3\eta } \right)}}{{3{{\left( {\Omega {T^4}} \right)}^{1/3}}}} + \frac{{{{\left( {{T^4}\Omega } \right)}^{1/3}}}}{{6{{\left( {2\pi } \right)}^{2/3}}{T^2}}},
\end{equation}
where we note that $\Omega  = 32{\pi ^2}{T^2}{r_c}\left( {r_c^2 + 9\eta } \right) - 27{r_c} + 3\sqrt 3 \left\{ {1024{\pi ^4}{T^4}{\eta ^3} + 64{\pi ^2}r_c^4{T^2}\left( {16{\pi ^2}{T^2}\eta  - 1} \right)} \right.$ ${\left. { + r_c^2\left[ {27 + 64\left. {{\pi ^2}{T^2}\eta \left( {32{\pi ^2}{T^2}\eta  - 9} \right)} \right]} \right.} \right\}^{\frac{1}{2}}}$. Combining the above relation with thermodynamic pressure~(\ref{eq12}), the equation of state $P$ is given in terms of  $T,{r_c},\eta$ as
\begin{align}
\label{eq14}
& P=   - \frac{1}{{4\pi r_c^2}} + \frac{{\sqrt {\frac{2}{3} - \frac{{8  {{\left( {2{\pi ^4}{T^4}} \right)}^{1/3}}\left( {r_c^2 - 3\eta } \right) + {\Omega ^{2/3}}}}{{6{r_c}{{\left( {4{\pi ^2}{T^2}\Omega } \right)}^{1/3}}}}} }}{{4\pi r_c^2}}
 \nonumber \\
&+ \frac{1}{{4r_c^2\sqrt {24{\pi ^2} - \frac{{24  {{\left( {4{\pi ^8}{T^2}} \right)}^{1/3}}\left( {r_c^2 - 3\eta } \right)}}{{{r_c}{\Omega ^{1/3}}}} - \frac{{3  {{\left( {2{\pi ^4}\Omega } \right)}^{1/3}}}}{{{r_c}{T^{2/3}}}}} }}
 \nonumber \\
 & + \frac{{{T^{2/3}}}}{{{{\left( {2\Omega } \right)}^{1/3}}{r_c}\sqrt {24{\pi ^{2/3}} - \frac{{24  {{\left( {4{\pi ^4}{T^2}} \right)}^{1/3}}\left( {{r^2} - 3\eta } \right)}}{{{r_c}{\Omega ^{1/3}}}} - \frac{{3{{\left( {2\Omega } \right)}^{1/3}}}}{{{r_c}{T^{2/3}}}}} }}
   \nonumber \\
&  - \frac{{{T^{2/3}}\eta }}{{2{{\left( {2\pi \Omega } \right)}^{1/3}}r_c^3\sqrt {\frac{2}{3} - \frac{{8  {{\left( {2{\pi ^4}{T^4}} \right)}^{1/3}}\left( {r_c^2 - 3\eta } \right) - {\Omega ^{2/3}}}}{{6{{\left( {4{\pi ^2}{T^2}\Omega } \right)}^{1/3}}{r_c}}}} }}.
\end{align}
This result indicates that the equation of state~(\ref{eq12})  depends not only on the temperature $T$  and the radius of the cavity  $r_c$, but also on the rainbow parameter $\eta$. Next, to obtain the critical points of this canonical ensemble system, it is necessary to solve the following equations:
\begin{equation}
\label{eq15}
{\left( {\frac{{\partial P}}{{\partial {r_c}}}} \right)_T} = {\left( {\frac{{{\partial ^2}P}}{{\partial r_c^2}}} \right)_T} = 0.
\end{equation}
In principle, one can obtain two relations for obtaining critical points of the canonical ensemble system by substituting equation of state~(\ref{eq14}) into Eq.~(\ref{eq15}). However, it is difficult to derive the critical quantities analytically, hence, we have to use a numerical method. For later convenience, we consider $\eta=1$, in which case the effect of RG becomes strong when the energy approaches $E_p$. The critical quantities becomes
\begin{align}
\label{eq16}
& {P^{{\rm{critical}}}} \approx 0.014,r_c^{{\rm{critical}}} \approx 1.137, {T^{{\rm{critical}}}} \approx 0.119,
 \nonumber \\
&\frac{{{P^{{\rm{critical}}}}r_c^{{\rm{critical}}}}}{{{T^{{\rm{critical}}}}}} = 0.135.
\end{align}
To check whether the obtained values are the ones in which phase transition takes place, we need to depict the  $P - {r_c}$, $T - {r_c}$ and $C - {r_c}$  diagrams.
\begin{figure}
\centering 
\includegraphics[width=.53\textwidth,origin=c,angle=0]{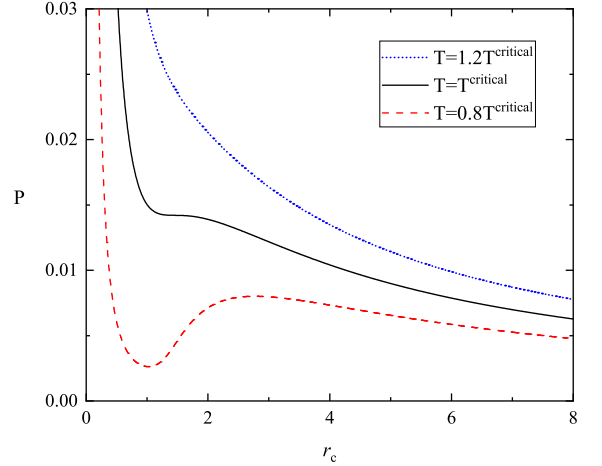}
\caption{\label{fig1} The pressure of the rainbow black hole versus the radius of the cavity for different temperatures with $\eta=1$.}
\end{figure}

According to Eq.~(\ref{eq14}), the corresponding  ``$P - {r_c}$" diagram is plotted in Fig.~\ref{fig1}. One can see that the temperature $T$  decreases from top to bottom. The isotherm (blue dotted curve) of the black hole system decreases monotonically with increasing  ${r_c}$ for  $T > {T^{{\rm{critical}}}}$, which looks like that of the ideal gas. The black solid curve corresponds to the critical isotherm $T = {T^{{\rm{critical}}}}$. When $T < {T^{{\rm{critical}}}}$, the red dashed isotherm has a significant oscillating part, which is reminiscent of the ``pressure-temperature'' relation of the Van der Waals gas or the ``pressure-specific volume" relation of the R-N AdS black hole \cite{cha10}.  Therefore, according to the $P - {r_c}$ plane, one can find that the black hole system exhibits a phase transition when the temperature is less than ${T^{{\rm{critical}}}}$.

\begin{figure}[htbp]
\centering
\subfigure[$\eta=1$]{
\begin{minipage}[b]{0.41\textwidth}
\includegraphics[width=1.2\textwidth]{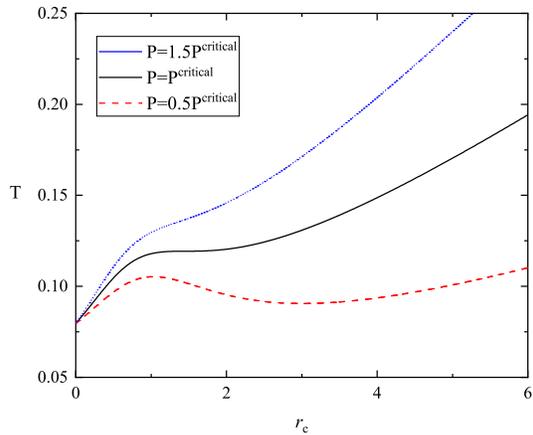}
\label{fig2-a}
\end{minipage}
}
\subfigure[$P=0.5{P^{{\rm{critcal}}}}$]{
\begin{minipage}[b]{0.41\textwidth}
\includegraphics[width=1.2\textwidth]{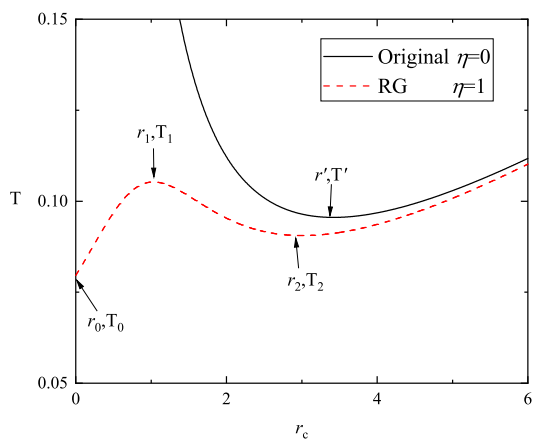}
\label{fig2-c}
\end{minipage}
}
\caption{(a) The temperature of the rainbow SC black hole surrounded by a cavity in the extended phase space for fixed rainbow parameter ($\eta=1$) and varying pressure. (b) The original rainbow SC black hole surrounded by a cavity in the extended phase space as a function of  $r_c$ with $P = 0.5{P^{{\rm{critcal}}}}$.}
\label{fig2}
\end{figure}

Next, by solving $r_c$ in terms of  $P$~(\ref{eq12}), one has ${r_H} = 4\Theta$, where  $\Theta  = \sqrt {2\pi P\left( {1 + 2P\pi r_c^2} \right)}$ $\left( {1 + 4P\pi r_c^2} \right)r_c^2 - 4P\pi r_c^3(1 + 2p\pi r_c^2)$. Putting the expression of ${r_H}$ into the expression of the temperature of the rainbow SC black hole in a cavity~(\ref{eq9}), one has
\begin{equation}
\label{eq18}
T\left( {{r_c},\eta ,P} \right) = \frac{1}{{4\pi \sqrt {\left( {{\Theta ^2} + \eta } \right)\left( {1 - \frac{\Theta }{r}} \right)} }}.
\end{equation}
Clearly, the temperature of the rainbow SC black hole surrounded by a cavity in the extended phase space is dependent on  $r_c$,  $P$ and $\eta$. According to Eq.~(\ref{eq18}), the temperature  $T$ as a function $r_c$ is depicted in Fig.~\ref{fig2}.

Fig.~\ref{fig2-a} shows that for  $\eta\neq 0$, the curves have an obvious undulating behavior when $P < {P^{{\rm{critical}}}}$, indicating that the system undergoes a phase transition in this case. As shown in Fig.~\ref{fig2-c}, the red dashed curve for $\eta=1$ represents to the original case at a large scale, which indicates that the effect of QG is negligible at that scale. However, as the radius decreases, the behavior of the modified temperature gradually deviates from the original case due to the effect of rainbow gravity. One can see that the temperature reduces to $T_2$ at $r_2$ and then increases to the peak $T_1$ at $r_1$. Interestingly, the temperature drops to a limited value $T_0$ as the radius becomes $r_0=0$ at the end of evolution. This turns out to be a ``massless remnant" \cite{cha54} corresponding to a finite temperature ${T_{{\rm{rem}}}} = T_0   = {1 \mathord{\left/ {\vphantom {1 {4\pi \sqrt \eta  }}} \right. \kern-\nulldelimiterspace} {4\pi \sqrt \eta  }}$. Moreover, this result can be interpreted as the black hole dissolving into particles of temperature following the argument in Ref.~\cite{cha55}.

Now, to study the thermodynamic stability and the phase structure of the black hole system, one should calculate the heat capacity, whose expression presents:
\begin{equation}
\label{eq19}
C = T\left( {\frac{{\partial S}}{{\partial T}}} \right).
\end{equation}
Substituting Eq.~(\ref{eq8}) and Eq.~(\ref{eq18}) into Eq.~(\ref{eq19}), one can obtain the heat capacity. However,  the expression is too long to express here, but we can  plot $C - {r_c}$ planes to illustrate the heat capacity of the rainbow SC black hole versus the radius $r_c$.

\begin{figure}[htbp]
\centering
\subfigure[$\eta=1$]{
\begin{minipage}[b]{0.41\textwidth}
\includegraphics[width=1.1\textwidth]{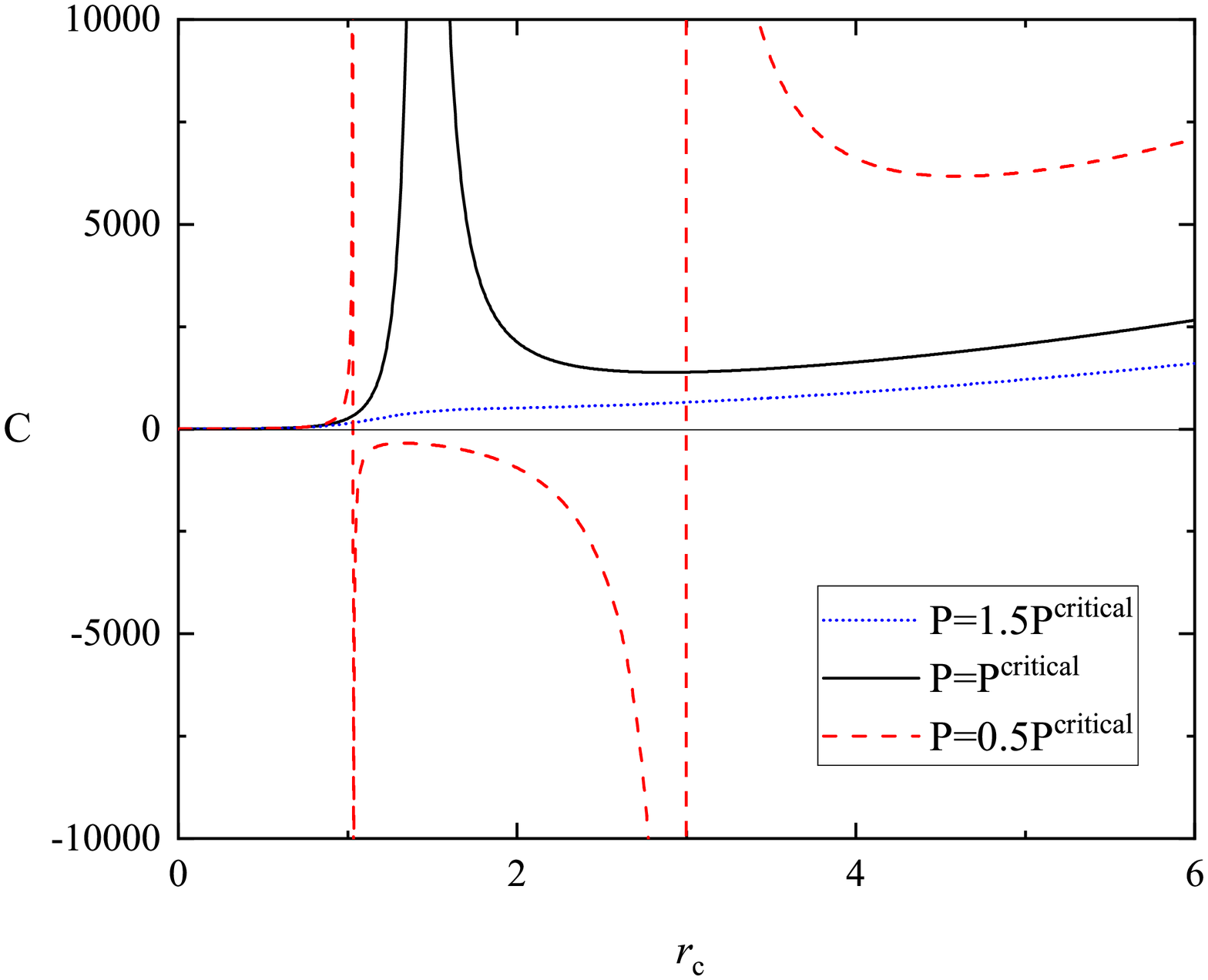}
\label{fig3-a}
\end{minipage}
}
\subfigure[$P=0.5{P^{{\rm{critcal}}}}$]{
\begin{minipage}[b]{0.41\textwidth}
\includegraphics[width=1.1\textwidth]{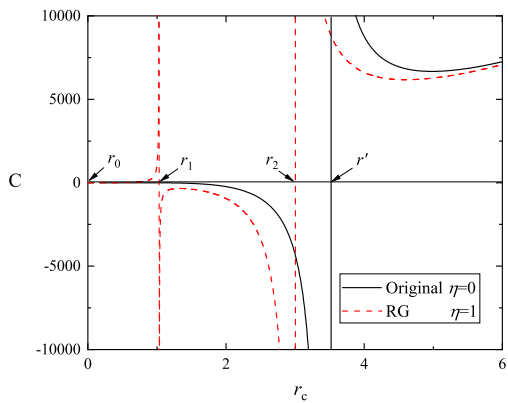}
\label{fig3-c}
\end{minipage}
}
\caption{(a) The heat capacity of the rainbow SC black hole surrounded by a cavity in the extended phase space for fixed rainbow parameter ($\eta=1$) and varying pressure. (b) The rainbow heat capacity of the SC black hole surrounded by a cavity in the extended phase space as a function of  $r_c$ with $P = 0.5{P^{{\rm{critcal}}}}$.}
\label{fig3}
\end{figure}

As shown in Fig.~\ref{fig3-a}, the pressure decreases from top to bottom. It is obvious that the rainbow heat capacity remains positive (black solid curve and blue dotted curve) for $P \ge {P^{{\rm{critical}}}}$, whereas the red dashed line for $P < {P^{{\rm{critical}}}}$ is distributed on both sides of the horizontal curve. Additionally, it is interesting to analyze how the rainbow effect changes the heat capacity of the SC black hole surrounded by a cavity in the extended phase space. From Fig.~\ref{fig3-c}, one can see the red dashed curve for the rainbow SC black hole case diverges at $r_1$ and $r_2$. It is well known that the positive and negative heat capacities determine the stability of black holes. Therefore, there exist two stable regions (${r_0} < r < {r_1}$ and $r>r_2$) and one unstable region (${r_1} < r < {r_2}$).

According to the above discussions, when $P$ or $T$ is less than the values obtained in Eq.~(\ref{eq16}), the system undergoes the phase transition. Besides, as shown in Fig.~\ref{fig2-c} and Fig.~\ref{fig3-c}, the rainbow SC black hole surrounded by a cavity in the extended phase space can be divided into three branches depending on their scales of  $r_c$. To gain an intuitive understanding, the range, state, heat capacity, and stability of the branches of the rainbow SC black hole surrounded by a cavity in the extended phase space are listed in Table~\ref{tab1}.
\begin{table}[htbp]
\centering
\caption{\label{tab1} Stability, radius,  state and heat capacity of the rainbow SC black hole surrounded by a cavity in the extended phase space.}
\begin{tabular}{c c c  c  c}
\hline
Branches       &Region                            &State           &Heat capacity            &Stability \\
\hline
1                       &${r_0} < r < {r_1}$       &SBH             &$C > 0$                         &stable\\
2                       &${r_1} < r < {r_2}$       &IBH              &$C < 0$                         &unstable\\
3                       &$r > {r_2}$                      &LBH             &$C > 0$                        &stable \\
\hline
\end{tabular}
\end{table}

From Table~\ref{tab1}, due to the effect of rainbow gravity, the SC case has three branches; it not only has a stable SBH and a stable LBH, but also has an additional unstable intermediate black hole (IBH) which never appears in the original SC black hole case. In addition, one may find that there is a remnant in the final stages of black hole evolution since the SBH of the rainbow case is stable.

\section{Phase transition of rainbow SC black hole enclosed by a cavity in extended phase space}
\label{sec4}
\begin{figure}[htbp]
\centering
\subfigure[$\eta=1$]{
\begin{minipage}[b]{0.41\textwidth}
\includegraphics[width=1.1\textwidth]{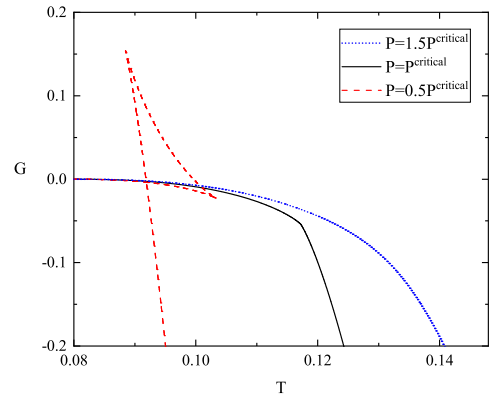}
\label{fig4-a}
\end{minipage}
}
\subfigure[$\eta=1$, $P = 0.5{P^{{\rm{critical}}}}$]{
\begin{minipage}[b]{0.41\textwidth}
\includegraphics[width=1.1\textwidth]{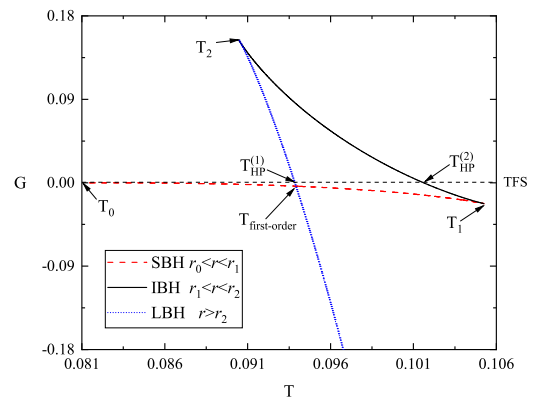}
\label{fig4-c}
\end{minipage}
}
\caption{(a)  Gibbs free energy of the rainbow SC black hole versus temperature for different pressures. (b) Gibbs free energy of the rainbow SC black hole as a function of temperature .}
\label{fig4}
\end{figure}

Finally, it is interesting to investigate the thermodynamic phase transition of the rainbow SC black hole surrounded by a cavity in the extended phase space. In the extended phase space, the Gibbs free energy of the SC black hole surrounded in a cavity can be expressed as follows \cite{cha56}
\begin{equation}
\label{eq20}
G = \mathcal{E} + PV - TS.
\end{equation}
Substituting Eq.~(\ref{eq8}), Eq.~(\ref{eq9}) into Eq.~(\ref{eq20}),  yields
\begin{align}
\label{eq21}
G & =  {r_c} \left( {1 - \sqrt {1 - \frac{{4\Theta }}{{{r_c}}}} } \right) - \frac{{r_c}}{3}\left( {1 - \frac{{4\Theta }}{{2r\sqrt {1 - \frac{{4\Theta }}{{{r_c}}}} }} - \sqrt {1 - \frac{{4\Theta }}{{{r_c}}}} } \right)
  \nonumber \\
& - \frac{{4\Theta \sqrt {{{\left( {4\Theta } \right)}^2} + \eta }  + \eta \ln \left( {4\Theta  + \sqrt {{{\left( {4\Theta } \right)}^2} + \eta } } \right)}}{{4\sqrt {1 - \frac{{4\Theta }}{{{r_c}}}} \sqrt {{{\left( {4\Theta } \right)}^2} + \eta } }}.
\end{align}
Now, using the Eq.~(\ref{eq21}) together with temperature~(\ref{eq18}), the ``$G - T$ diagrams'' for the SC black hole surrounded by a cavity in the extended phase space are displayed in Fig.~\ref{fig4}.

In Fig.~\ref{fig4-a}, by setting $\eta  = 1$, the $G$ of the rainbow SC black hole surrounded by a cavity in the extended phase  is depicted as a function of $T$ with various pressures $P = 1.5{P^{{\rm{critical}}}}$, $P = {P^{{\rm{critical}}}}$ and $P = 0.5{P^{{\rm{critical}}}}$. Obviously, the phase structure of the rainbow SC black hole surrounded by a cavity in the extended phase space is similar to that of the R-N AdS black hole. The black solid curve for $P < {P^{{\rm{critical}}}}$ shows that a swallow tail structure corresponds to a two-phase coexistence state phase transition in the canonical ensemble, whereas the system only has one thermally stable phase with $P \ge {P^{{\rm{critical}}}}$. These results are consistent with the profiles of  ``$T - {r_c}$" in Fig.~\ref{fig2} and ``$C - {r_c}$" in Fig.~\ref{fig3}.

Next, to reveal the influence of the rainbow's effect on the phase transition, we display Fig.~\ref{fig4-c}. As shown in Fig.~\ref{fig4-c}, the horizontal dashed line is the free energy of thermal flat spacetime (TFS), which is a classical solution of the canonical ensemble. One can see that the system only has SBH for $\left( {{T_0},{T_2}} \right)$ or LBH for $T > {T_1}$, while the three branches of the rainbow SC black hole coexist for $\left( {{T_2},{T_1}} \right)$. The swallow tail leads to one first-order phase transition at the inflection point ${T_{{\rm{first - order}}}}$  that occurs at Van der Waals fluid/AdS charged black hole system in the extended phase space, and never at the original SC black hole case \cite{cha10}. However, one may see that the Gibbs free energies of the SBH and the LBH at the two ends of the first-order phase transition are both less than zero, which also appears in the high-dimensional  AdS black hole in massive gravity \cite{chy1}. Those imply that the effect of RG plays a role of charge in the evolution of black holes, and can effectively reduce the free energy of the black hole  thermodynamic  ensemble. In addition, by comparing the Gibbs free energies of different phases, it can be observed that ${G^{{\rm{LBH}}}}$  is lower than ${G^{{\rm{IBH}}}}$,  ${G^{{\rm{TFS}}}}$ and  ${G^{{\rm{SBH}}}}$ for $T > {T_{{\rm{first - order}}}}$, which implies that the IBH, TFS and IBH would decay into a stable LBH. However, as the temperature decreases, the relationship of the Gibbs free energy changes to ${G^{{\rm{SBH}}}} < {G^{{\rm{LBH}}}} < {G^{{\rm{TFS}}}} < {G^{{\rm{IBH}}}}$ for $T_{{\text{HP}}}^{(1)} < T < {T_{{\text{first - order}}}}$, and ${G^{{\rm{SBH}}}} < {G^{{\rm{TFS}}}} < {G^{{\rm{LBH}}}} < {G^{{\rm{IBH}}}}$ for $ T < T_{{\text{HP}}}^{(1)}$. Hence, the other three phases would decay into the stable SBH in this region. Even more interesting is that the black solid line for the IBH and blue dotted line for the LBH intersect with the horizontal line at points $T_{{\rm{HP}}}^{\left( 1 \right)}$ and  $T_{{\rm{HP}}}^{\left( 2 \right)}$, which implies that the rainbow SC black hole system contains two Hawking-Page-like phase transitions. From Fig.~\ref{fig4-c} and Fig.~\ref{fig3-c}, it can be seen that the unstable IBH cannot exist for a long time. Therefore, the $T_{{\rm{HP}}}^{\left( 2 \right)}$ is in a process of unstable phase transition and $T_{{\rm{HP}}}^{\left( 1 \right)}$ exist as a metastable phase transition.

\section{Conclusion and discussion}
\label{sec5}
In this paper, by defining a new thermodynamic volume ${{V = 4\pi r_c^3} \mathord{\left/ {\vphantom {{V = 4\pi r_c^3} 3}} \right. \kern-\nulldelimiterspace} 3}$  and its conjugate thermodynamic pressure $P =  - \left( {{{\partial {\cal E}} \mathord{\left/ {\vphantom {{\partial {\cal E}} {\partial V}}} \right. \kern-\nulldelimiterspace} {\partial V}}} \right)$, we investigated the thermodynamic properties, critical behavior and phase structure of the rainbow SC black hole enclosed by a cavity in extended phase space.  We summarize our results as follows:

\begin{enumerate}
\item We restored the $P-V$ term and calculated the critical quantities of the rainbow SC black hole surrounded by a cavity in the extended phase space. The critical pressure ${P^{{\rm{critical}}}}$  and critical temperature ${T^{{\rm{critical}}}}$ decrease with $\eta$, while the critical radius ${r^{{\rm{critical}}}}$ and critical ratio decrease with $\eta$.

\item When $T < {T^{{\rm{critical}}}}$, the behavior of $P - {r_c}$ criticality from Fig.~\ref{fig1} is reminiscent of Van der Waals gas, which indicates that the system undergoes a phase transition. For $P < {P^{{\rm{critical}}}}$, the $T - {r_c}$ plane and $C - {r_c}$ plane also confirm that the canonical ensemble has a phase transition.

\item  Rainbow gravity leads to a massless remnant with limited temperature ${T_{{\rm{rem}}}} = {1 \mathord{\left/ {\vphantom {1 {4\pi \sqrt \eta  }}} \right. \kern-\nulldelimiterspace} {4\pi \sqrt \eta  }}$ in the final stages of black hole evolution. This result can be interpreted as a black hole dissolves at a finite temperature into a particle that can store information, thereby avoiding the black hole information paradox.

\item In the extended phase space, due to the effect of rainbow gravity,  the SC black hole in a cavity has two stable regions and one unstable region. Therefore, the black hole system is naturally divided into three branches, that is, a stable LBH, a stable SBH, and an unstable IBH that never appears in the original case and AdS counterpart case. On the phase transition side, the system exhibits two Hawking-Page-like phase transitions and one first-order phase transition, which is very different from the original case.

\item It is worth noting that, in order to ensure that the wall of the cavity can effectively reflect the radiation of rainbow SC black hole, the radius of the cavity and the event horizon of SC black hole satisfy the relationship $0\leq{{{r_H}} \mathord{\left/ {\vphantom {{{r_H}} {{r_c}}}} \right.  \kern-\nulldelimiterspace} {{r_c}}}\leq1$, hence, when ${r_c} \to \infty$, the rainbow SC black hole would loss the reflecting surface, which makes it impossible to study the phase behavior of the system. On this base, if one further consider that $\eta  \to 0$, the thermodynamic quantities of rainbow SC black hole (i.e, the Hawking temperature, entropy and heat capacity) would reduce the case of the original SC black hole.
\end{enumerate}

In Refs.~\cite{cha29,cha30}, the authors demonstrated that the thermodynamic properties, critical behavior and phase structure of black holes surrounded by a cavity in the extended phase space are almost the same as their AdS counterparts. However, due to the effect of RG, we found that the thermodynamic behavior and phase transition of the rainbow SC black hole surrounded by a cavity in the extended phase space is quite similar to that of the R-N AdS black hole. This indicates that the effect of quantum gravity can significantly change the thermodynamic properties and phase transition of black holes.  In addition, Garattini and Saridakis~\cite{chy2} have pointed out that the  rainbow gravity corresponds to the  H\v{o}rava-Lifshitz gravity,  hence, the investigation of the thermodynamic properties of rainbow SC black hole could open a new window for further understanding of H\v{o}rava-Lifshitz gravity. It is believed that the relevant research would be very interesting, and we will discuss it in detail in future works.

\end{document}